\documentclass{aa}
\usepackage{graphicx,multirow}
\usepackage{amsmath}
\usepackage{natbib}
\bibpunct{(}{)}{;}{a}{}{,}   

\begin{document}

\title{Observational constraints on energetic particle diffusion in young SNRs: amplified magnetic field and maximum energy}

\author{E. Parizot\inst{1} \and A. Marcowith\inst{2} \and J. Ballet \inst{3} \and  Y. A. Gallant \inst{4}}

\institute{IPN Orsay, CNRS/Universit\'e Paris-Sud, 91406 Orsay Cedex, France \and Centre d'\'Etude Spatiale des Rayonnements, 9 av. du Colonel Roche, 31028 Toulouse Cedex 4, France \and DSM/DAPNIA/SAp, CEA Saclay, 91191 Gif sur Yvette Cedex, France \and LPTA, CNRS/Universit\'e Montpellier II, 34095 Montpellier Cedex 5, France}

\date{Received ??; Accepted ??}

\abstract{Constraints on the diffusion and acceleration parameters in five young supernova remnants (SNRs) are derived from the observed thickness of their X-ray rims, as limited by the synchrotron losses of the highest energy electrons, assuming uniform and isotropic turbulence. From a joint study of the electrons diffusion and advection in the downstream medium of the shock, it is shown that the magnetic field must be amplified up to values between 250 and 500~$\mu$G in the case of Cas A, Kepler, and Tycho, or $\sim 100\,\mu$G in the case of SN 1006 and G347.3-0.5. The diffusion coefficient at the highest electron energy can also be derived from the data, by relating the X-ray energy cutoff to the acceleration timescale. Values typically between 1 and 10 times the Bohm diffusion coefficient are found to be required. We also find interesting constraints on the energy dependence of the diffusion coefficient, by requiring that the diffusion coefficient at the maximum proton energy be not smaller than the Bohm value in the amplified field. This favours diffusion regime between the Kraichnan and the Bohm regime, and rejects turbulence spectrum indices larger than $\simeq 3/2$. Finally, the maximum energy of the accelerated particles is found to lay between $10^{13}$ and $5\,10^{13}$~eV for electrons, and around $Z\times 8\,10^{14}$~eV at most for nuclei (or $\sim 2.5$ times less if a Bohm diffusion regime is assumed), roughly independently of the compression ratio assumed at the shock. Even by taking advantage of the uncertainties on the measured parameters, it appears very difficult for the considered SNRs in their current stage of evolution to produce protons up to the knee of the cosmic-ray spectrum, at $\sim 3\,10^{15}$~eV, and essentially impossible to accelerate Fe nuclei up to either the ankle at $\sim 3\times 10^{18}$~eV or the second knee at $\sim 5\times 10^{17}$~eV.
}

\titlerunning{Observational constraints on energetic particle diffusion in young SNRs}

\maketitle

\section{Introduction}
\label{sec:intro}

High-angular resolution X-ray observations of young supernova remnants (SNRs) show very thin rims of emission, associated with the forward shock of the supernova (SN) expanding in the interstellar medium (ISM).  This emission is most probably related to the synchrotron emission of high-energy electrons \citep[and references therein]{Ballet05} accelerated at the shock by the well-known diffusive shock acceleration (DSA) mechanism, which requires the scattering of electrons back and forth through the velocity discontinuity associated with the shock.  While this mechanism is well understood qualitatively and has been extensively confronted with multi-wavelength observational data, as well as direct measurement in interplanetary shocks, very little is known about one of its most fundamental ingredients: the particle diffusion coefficient.

The diffusion mechanism is related to the deflection of charged particles in the ambient magnetic field and the resonant interaction with MHD waves.  The value and structure of the inhomogeneous magnetic field is thus crucial to determine $D(E)$. In the absence of a fully predictive theory of wave generation and particle diffusion inside SNRs, it is generally assumed that the diffusion coefficient is close to the so-called Bohm limit \emph{at all energies}, defined by $D_{\mathrm{B}}(E) = r_{\mathrm{L}}v/3$, where $r_{\mathrm{L}} = p/qB$ is the Larmor radius of the particle of momentum $p$ in the field $B$, and $q$ and $v$ are its charge and velocity.  This value of $D$ corresponds to a mean free path of the charged particles equal to the Larmor radius, which is thought to be the lowest possible value for isotropic turbulence. However, particles of different energies resonate with field fluctuations at different scales, and the magnetic field intensity to be used to calculate $r_{\mathrm{L}}$ (and thus $D_{\mathrm{B}}$) should be an effective value at the resonant scale ($\lambda \simeq r_{\mathrm{L}}$). The diffusion coefficient therefore depends in principle on the power spectrum of the magnetic field. In particular, $D(E) \propto E^{1/3}$ is expected in the quasi-linear regime in the case of a Kolmogorov-like spectrum of turbulence. Recent numerical analyses of particle diffusion in isotropic turbulent magnetic fields have confirmed analytical results in their range of validity and shown that the Bohm regime does not generally hold \citep{Cetal02,Parizot04,CanRou04}, even in the strong turbulence limit, and can be approached only at the critical energy where $r_{\mathrm{L}} = \lambda_{\mathrm{c}}$, the coherence length of the field. The question thus remains whether the Bohm approximation is relevant to the case of DSA in SNRs.

\begin{table*}
\begin{center}
\begin{tabular}{c|c|c|c|c|c}
Name & Distance & Shock speed & Projected width & Age & Cut-off energy \\
\hline
Cas A & 3.4 kpc $^1$ & 5200 km/s $^2$ & 0.05 pc (3\arcsec) $^3$ & 320 yr ? $^4$ & 1200 eV $^3$ (whole SNR) \\
Kepler & 4.8 kpc $^5$ & 5400 km/s $^6$ & 0.07 pc (3\arcsec) $^7$ & 400 yr & 900 eV $^8$ \\
Tycho & 2.3 kpc $^9$ & 4600 km/s $^{10}$ & 0.05 pc (4\arcsec) $^7$ & 430 yr & 290 eV $^{11}$ \\
SN 1006 & 2.2 kpc $^{12}$ & 2900 km/s $^{13}$ & 0.2 pc (20\arcsec) $^{14}$ & 1000 yr & 3000 eV $^{15}$ \\
G347.3$-$0.5 & 1.3 kpc $^{16}$ & 4000 km/s ? $^{17}$ & 0.25 pc (40\arcsec) $^{18}$ & 1620 yr ? $^{19}$ & 2600 eV $^{18}$ \\
\end{tabular}
\end{center}
\caption{Characteristics of the non-thermal emission behind the blast wave in young SNRs (see Ballet, 2005, for further detail). The numerical values are taken (or directly inferred) from: $^1$ \citet{rh95}, $^2$ \citet{vb98}, 
$^3$ \citet{vl03} from a fit to the spectrum of the whole SNR, 
$^4$ \citet{as80}, $^5$ \citet{rg99}, $^6$ \citet{hu99}+distance, 
$^7$ Decourchelle (2003) (private communication) , $^8$ \citet{cd04} in the southeast, 
$^9$ \citet{sk91}, $^{10}$ \citet{hu00},
$^{11}$ \citet{hd02}, $^{12}$ \citet{wg03},
$^{13}$ \citet{gw02} in the northwest, 
$^{14}$ \citet{by03} in the northeast, $^{15}$ \citet{Retal04} in the northeast,
$^{16}$ \citet{ca04}, $^{17}$ from $v = \lambda r/t$ with the expansion parameter $\lambda = 2/3$, 
$^{18}$ \citet{ls03} in the northwest, $^{19}$ \citet{wq97}.
}
\label{nt_snrs}
\end{table*}

A striking property of the DSA mechanism, however, is that in the steady-state regime and the so-called test-particle approximation, the global shape of the energetic particle (EP) energy spectrum does not depend on the exact value of the diffusion coefficient, nor even on its variation with both energy and space. Many studies could thus be made without precise knowledge of the diffusion mechanism.  In particular, the `universal' $E^{-2}$ spectrum could be obtained in the test-particle limit, and some insights were gained into the non-linear regime where the EPs influence the shock structure (see e.g. the review by Jones and Ellison, 1991). The acceleration rate, however, is directly related to the spatial diffusion coefficient: the smaller this parameter, $D(E)$, the smaller the acceleration time, $\tau_{\mathrm{acc}}$.  To some factor of order unity, $\tau_{\mathrm{acc}} \simeq D(E)/V_{\mathrm{sh}}^{2}$, where $V_{\mathrm{sh}}$ is the shock velocity.  For this reason, the value (and energy dependence) of $D(E)$ governs the highest energy which can be reached in a SNR.

Another aspect of diffusion, of course, is the transport of particles in space!  By allowing one to draw maps of the non thermal emission, high-resolution observations of SNRs can provide information about the EP transport properties \citep{vdS04}.  The other ingredients governing particle transport are advection and energy losses.  In the case of electrons, the latter are also related to the ambient magnetic field, and thus are not independent of the diffusion coefficient.

In this paper, we extend previous work on specific objects and transport models \citep{Betal03,BV04,Rey04,V04,Yetal04,Vetal05}, investigating the constraints that high-angular resolution observations can set on two key parameters: the magnetic field and the diffusion coefficient. We discuss the interplay between the various phenomena governing electron transport in SNRs, assuming that the turbulence is isotropic: investigation of anisotropic effects are beyond the scope of the paper.  We make the further assumption that the X-ray rims thickness is limited by the radiative losses and leave for future work the discussion of the impact of the turbulence relaxation downstream of the forward shock on the X-ray filaments morphology (see Pohl et al. 2005). We apply our analyses to five young SNRs observed by Chandra and discuss the resulting maximum energy of electrons and protons in SNRs. For all numerical applications, we assume the reference values summarized in Table~\ref{nt_snrs}.

\section{Electron transport at the SNR shock}
\label{sec:constraints}

\subsection{The size of the X-ray rims}

In the DSA theory, energetic electrons are assumed to be scattered by magnetic inhomogeneities (or MHD waves) hosted by the underlying plasma, and diffuse roughly isotropically in the plasma rest frame. With respect to the shock front, two processes compete: i) global advection with the fluid, towards the shock upstream and away from it downstream, and ii) spatial diffusion which allows a fraction of the particles advected downstream to nevertheless cross back the shock and gain energy through the standard first order Fermi mechanism. Diffusion also makes the upstream particles explore a region of size $\sim D/V_{\mathrm{sh}}$ ahead of the shock, where $V_{\mathrm{sh}}$ is the shock velocity. In addition, the highest energy electrons suffer from efficient energy losses due to synchrotron emission in the local magnetic field, which limit their maximum energy and lead to an assumed exponential cut-off in their spectrum: $f_{\mathrm{e}}(E) \propto E^{-\gamma}\times\exp(-E/E_{\mathrm{e,max}})$. Because of these losses, the highest energy electrons cannot travel far from the acceleration region and their emission must be limited to very thin regions just behind the shock, seen as sharp and narrow rims in X-ray.

The \emph{observed} thickness of the rims, $\Delta R_{\mathrm{obs}}$, is related to their actual size by a \emph{projection factor}, $P$, which depends on the geometry of emission region. From simple arguments based on the transport equation of the accelerated electrons (see below), one can infer that the electron distribution downstream has an exponential profile: $f_{\mathrm{e}}(x) \propto \exp(-x/\Delta R_{\mathrm{rim}})$, where $x$ is the radial coordinate measured from the shock front, with increasing values downstream, and $\Delta R_{\mathrm{rim}}$ defines the characteristic scale of the electron distribution. Note that $\Delta R_{\mathrm{rim}}$ depends on the electron energy. Here, we always refer to the observed rims at $E_{\gamma} = 5$~keV, corresponding to electrons of energy $E \equiv E_{\mathrm{e,obs}}$.

In the ideal case of a spherical shock, an exponential emission profile translates into a projected thickness $\Delta R_{\mathrm{obs}} \simeq 4.6 \Delta R_{\mathrm{rim}}$, where $\Delta R_{\mathrm{obs}}$ is defined as the size of the region where the observed brightness is larger than half of the maximum (Ballet, 2005). This was obtained by numerical calculation in the limit $\Delta R_{\mathrm{obs}}\ll R_{\mathrm{sh}}$, the SNR shock radius. Note that we neglected the contribution of the upstream emission, which should indeed be much thinner or even inexistent at the X-ray energies considered, because of the lower magnetic field upstream (see below). In the following, we adopt the above value of the projection factor, $P = 4.6$, and use the reduced projection factor $\bar{P} = P/4.6$ to track how the results depend on the assumed geometry.

\subsection{Synchrotron loss time}

Electrons of energy $E = \gamma m_{\mathrm{e}}c^2$ propagating in a magnetic field $B$ with a pitch angle $\alpha$ emit a synchrotron radiation at a typical energy of
\begin{equation}
E_{\gamma} \simeq \xi\frac{3h}{4\pi\beta}\gamma^{2} \frac{eB}{m_{\mathrm{e}}}\sin\alpha,
\label{eq:Egamma}
\end{equation}
where $\xi \simeq 0.29$ (e.g. Longair, 1994). This can be inverted to give the typical energy of an electron emitting photons at energy $E_{\gamma}\equiv E_{\gamma,\mathrm{keV}}$~keV. We shall assume here an approximate ``mono-energetic'' (i.e. one-to-one) conversion, as given by:
\begin{equation}
E_{\mathrm{e}} \simeq (22\,\mathrm{TeV})\times B_{100}^{-1/2}\times E_{\gamma,\mathrm{keV}}^{1/2},
\label{eq:EEgamma}
\end{equation}
where $B_{100}$ is the magnetic field in units of $100\,\mu\mathrm{G}$ (The impact of this simplification on our numerical results is negligible at the level of precision of the present approach). For the observed X-ray energy, $E_{\gamma} = 5$~keV, we thus have:
\begin{equation}
E_{\mathrm{e,obs}} \simeq (49\,\mathrm{TeV})\times B_{100}^{-1/2}.
\label{eq:EEgammaObs}
\end{equation}

When the electron leaves the acceleration process (i.e. it will not cross the shock front again), it cannot radiate at energy $E_{\gamma}$ for a time longer than the \emph{synchrotron energy loss time}, $\tau_{\mathrm{syn}}$, due to the synchrotron radiation itself (note that synchrotron losses are usually expected to be larger than inverse Compton losses, because of the large magnetic field). The corresponding energy loss rate is written $(\mathrm{d}E/\mathrm{d}t)_{\mathrm{syn}} \simeq -\frac{4}{3}\sigma_{\mathrm{T}}c \times \epsilon_{\mathrm{magn}} \times\beta^{2}\gamma^{2}$, where $\epsilon_{\mathrm{magn}} = B^{2}/2\mu_{0}$ is the magnetic energy density, $\sigma_{\mathrm{T}}$ is the Thomson cross section and $\beta = v/c \simeq 1$ in the case of interest here. Thus, noting $E_{\mathrm{TeV}}$ the electron energy in units of $10^{12}$~eV:
\begin{equation}
\tau_{\mathrm{syn}}\equiv\frac{E}{(\mathrm{d}E/\mathrm{d}t)_{\mathrm{syn}}} \simeq (1.25\times 10^{3}\,\mathrm{yr}) \times E_{\mathrm{TeV}}^{-1} \times B_{100}^{-2},
\label{eq:tauSynchE}
\end{equation}
or, at the energy corresponding to the observed rims:
\begin{equation}
\tau_{\mathrm{syn}}(E_{\mathrm{e,obs}}) \simeq (25\,\mathrm{yr})\times B_{100}^{-3/2}.
\label{eq:tauSynchEObs}
\end{equation}

\subsection{Advection downstream}
\label{sec:advectionDownstream}

During the time $\tau_{\mathrm{syn}}$, the downstream electrons are advected away from the shock at a velocity $V_{\mathrm{d}} = V_{\mathrm{sh}}/r$, where $r$ is the shock compression ratio ($r = 4$ for an unmodified strong shock, but it can be larger if non-linear effects are important, e.g. Ellison et al., 2004). The size of the advection region, where the electrons emit at 5~keV, is thus simply given by $\Delta R_{\mathrm{adv}} = V_{\mathrm{d}}\times \tau_{\mathrm{syn}}(E_{\mathrm{e,obs}})$. It can also be shown more formally that this is indeed the scale of the rims downstream, $\Delta R_{\mathrm{rim}}$, obtained when neglecting electron diffusion. The standard transport equation at a parallel shock is written (VÂ\"olk et al., 1981):
\begin{equation}
\frac{\partial f}{\partial t} + V\frac{\partial f}{\partial x} - \frac{p}{3}\frac{\partial f}{\partial p} \frac{\partial V}{\partial x} =  \frac{\partial}{\partial x}(D\frac{\partial f}{\partial x}) - \frac{f}{\tau_{syn}} + q,
\label{eq:transportEquation}
\end{equation}
where the last term, $q$, is an unspecified source term, and the loss term on the right hand side has been given a simplified form, corresponding to random catastrophic losses after a lifetime $\tau_{\mathrm{syn}}$. The stationary version of Eq.~(\ref{eq:transportEquation}) in the downstream region and without diffusion is simply $V_{\mathrm{d}}(\partial f/\partial x) = -f/\tau_{\mathrm{syn}}$, with the obvious solution $f_{\mathrm{e}}(x)\propto \exp(-x/V_{\mathrm{d}}\tau_{\mathrm{syn}})$.

Taking $\tau_{\mathrm{syn}}$ from Eq.~(\ref{eq:tauSynchEObs}), one obtains:
\begin{equation}
\Delta R_{\mathrm{adv}} \simeq (6.5\,10^{-3}\,\mathrm{pc})\times \frac{4}{r} \times V_{\mathrm{sh},3} \times B_{100}^{-3/2},
\label{eq:DRadvNum}
\end{equation}
where $V_{\mathrm{sh},3}$ is the shock velocity in units of $10^{3}$~km/s.

Even though the actual size of the rims, $\Delta R_{\mathrm{rim}}$, may be larger than $\Delta R_{\mathrm{adv}}$, because of electron diffusion (see below), the constraint $\Delta R_{\mathrm{adv}} \times P \le \Delta R_{\mathrm{obs}}$ is a simple consequence of the interpretation of the rims considered here. This straightforwardly translates into an upper limit on the synchrotron loss time and a lower limit on the magnetic field, $B_{\mathrm{d}}$, in the downstream region. Writing $\Delta R_{\mathrm{obs}} \equiv R_{\mathrm{obs},-2}\times 10^{-2}$~pc, one obtains:
\begin{equation}
\tau_{\mathrm{syn}} \le \frac{\Delta R_{\mathrm{obs}}}{V_{\mathrm{d}}} \simeq (8.5\,\mathrm{yr}) \times \frac{r}{4}\times \bar{P}^{-1} \times V_{\mathrm{sh},3}^{-1} \times \Delta R_{\mathrm{obs},-2},
\label{eq:tauSynUpperLimit}
\end{equation}
where $\bar{P} = P/4.6$ is the reduced projection factor, and
\begin{equation}
B_{\mathrm{d}} \ge (210\,\mu\mathrm{G}) \times (4\bar{P}/r)^{2/3}
\times V_{\mathrm{sh},3}^{2/3} \times \Delta R_{\mathrm{obs},-2}^{-2/3}.
\label{eq:BInfAdv}
\end{equation}

With the parameters in Table~\ref{nt_snrs}, Eq.~(\ref{eq:tauSynUpperLimit}) tells us that $\tau_{\mathrm{syn}}$ is typically smaller than 3\% of the age of the SNR, $t_{\mathrm{SNR}}$, except for SN 1006, where the upper limit is 6\% (see Table~\ref{tab:someLimits}). This is of course consistent with the assumption that the maximum electron energy is not limited by $t_{\mathrm{SNR}}$, and it also justifies that we use throughout the \emph{current} value of $V_{\mathrm{sh}}$.

The lower limit on the downstream magnetic field will be made more precise below. Here, we simply use Eq.~(\ref{eq:BInfAdv}) to derive a general limit related to the ``advection thickness'' of the rims, which we note $B_{\mathrm{adv}}$. For Cas A, Kepler, Tycho, SN 1006 and G347.3 + 0.5, one obtains respectively (see also Table~\ref{tab:someLimits}): 210, 170, 200, 57 and 61~$\mu$G (for $r = 4$). This does not seem compatible with the mere shock compression of the upstream magnetic field, unless the latter is significantly higher than in the average ISM and the compression ratio is of the order of 7--10, as could result from non-linear CR shock modification \citep{Ell04}. We thus argue that magnetic field amplification must occur around the shocks of young SNRs (see also below, and V\"olk et al., 2005, and references therein).

\subsection{Diffusion downstream}
\label{sec:diffDown}

In addition to their advection with the downstream plasma, the energetic electrons diffuse in the local turbulent magnetic field, and if their effective diffusion velocity away from the shock is larger than $V_{\mathrm{d}}$, they can fill regions larger than $\Delta R_{\mathrm{adv}}$. Writing again the stationary version of the transport equation~(\ref{eq:transportEquation}), but keeping the (uniform) diffusion term instead of the advection term, we obtain $\partial^{2}f/\partial x^{2} = f/D\tau_{\mathrm{syn}}$, and the solution $f_{\mathrm{e}}(x)\propto \exp(-x/\Delta R_{\mathrm{diff}})$, where $\Delta R_{\mathrm{diff}} = \sqrt{D\tau_{\mathrm{syn}}}$ thus gives the diffusive scale of the rims.

It was found above that the local magnetic field must have been amplified, therefore it is legitimate to assume $\delta B \gg B_{0}$, where $B_{0}$ is the regular field, and we also assume a roughly isotropic turbulence. In this case, even though the value of the diffusion coefficient, $D(E)$, is generally not known, it cannot be smaller than the so-called Bohm value, $D_{\mathrm{B}} = r_{\mathrm{L}}v/3$, as recalled in Sect.~\ref{sec:intro}:
\begin{equation}
D_{\mathrm{B}}(E) \simeq (3.3\times 10^{23}\,\mathrm{cm}^{2}\mathrm{s}^{-1}) \times E_{\mathrm{TeV}}\times B_{100}^{-1}.
\label{eq:DBohm}
\end{equation}
With the help of Eq.~(\ref{eq:tauSynchE}), one thus obtains:
\begin{equation}
\Delta R_{\mathrm{diff}} \ge (3.7\,10^{-2}\,\mathrm{pc})\times B_{100}^{-3/2},
\label{eq:DRdiffNum}
\end{equation}
independently of the electron energy (as a result of the Bohm scaling). Just as in the previous section, we now obtain a lower limit on the downstream magnetic field by imposing the condition $\Delta R_{\mathrm{diff}}\times P \le \Delta R_{\mathrm{obs}}$. This reads:
\begin{equation}
B_{\mathrm{d}} \ga (660\,\mu\mathrm{G})\times \bar{P}^{2/3} \times \Delta R_{\mathrm{obs},-2}^{-2/3}.
\label{eq:BFromDbohmRObs}
\end{equation}

This result is interesting since it provides a lower limit on the value of $B_{\mathrm{d}}$ that can be derived directly from the observed thickness of the synchrotron rims, independently of the shock velocity. Moreover, since it was obtained with the Bohm diffusion coefficient, any other assumption would lead to a larger diffusion scale, and thus thicker X-ray rims. The Bohm diffusion coefficient therefore provides the weakest possible constraint on the downstream magnetic field using the ``diffusion thickness'' of the rims, $B_{\mathrm{diff}}$, so that Eq.~(\ref{eq:BFromDbohmRObs}) is quite robust. Applying it to Cas A, Kepler, Tycho, SN 1006 and G347.3 + 0.5, with the values given in Table~\ref{nt_snrs}, one obtains, respectively: 230, 180, 230, 90 and 77~$\mu$G (cf. Table~\ref{tab:someLimits}). These lower limits on $B_{\mathrm{d}}$ are remarkably close to those obtained with $\Delta R_{\mathrm{adv}}$ and confirm that a strong amplification of the pre-existing field is required (Berezhko \& V\"olk, 2004, and V\"olk et al., 2005).

\begin{table}
\begin{center}
\begin{tabular}{c|c|c|c|c}
SNR & $\tau_{\mathrm{syn}}^{\mathrm{max}}/t_{\mathrm{SNR}}$ & $B_{\mathrm{adv}}$ & $B_{\mathrm{diff}}$ & \multirow{2}*{$\rho_{\mathrm{B}}=\frac{\Delta R_{\mathrm{diff}}^{(\mathrm{B})}}{\Delta R_{\mathrm{adv}}}$} \\
name & ($\times 4\bar{P}/r$) & ($\mu$G) & ($\mu$G) & \\
\hline
Cas A & $< 2.6$\%  & 210 & 230 & 1.1 \\
Kepler & $< 2.8$\% & 170 & 180 & 1.1 \\
Tycho & $< 2.1$\% & 200 & 230 & 1.2 \\
SN 1006 & $< 5.9$\% & 57 & 90 & 2.0 \\
G347.3$-$0.5 & $< 3.3$\% & 61 & 77 & 1.4 \\
\end{tabular}
\end{center}
\caption{Constraints on the synchrotron loss time and the downstream magnetic field for the five young SNRs whose observed parameters are given in Table~\ref{nt_snrs}. In columns 3 and 4, the constraints are lower limits derived from consideration of advection and diffusion only, respectively. To derive $B_{\mathrm{diff}}$ we assume a Bohm diffusion coefficient, which leads to the less constraining result (to the lowest magnetic field). In column 5, the superscript ``B'' indicates that the Bohm diffusion coefficient has also been assumed.}
\label{tab:someLimits}
\end{table}

\subsection{Diffusion vs. advection}

While diffusion and advection have been considered separately above, it is interesting to note that the corresponding thicknesses of the rims produced by the highest energy electrons accelerated at the shock should be of the same order of magnitude. Indeed, since $\Delta R_{\mathrm{diff}} = \sqrt{D\tau_{\mathrm{syn}}}$ and $\Delta R_{\mathrm{adv}} = V_{\mathrm{d}}\tau_{\mathrm{syn}}$, the condition $\Delta R_{\mathrm{adv}} = \Delta R_{\mathrm{diff}}$ is equivalent to $\tau_{\mathrm{syn}} = D/V_{\mathrm{d}}^2$. Now $D/V_{\mathrm{d}}^2$ gives the order of magnitude of the acceleration timescale, $\tau_{\mathrm{acc}}$, so that $\Delta R_{\mathrm{adv}}$ and $\Delta R_{\mathrm{diff}}$ are found to be comparable if and only if $\tau_{\mathrm{acc}} \simeq \tau_{\mathrm{syn}}$. This characterizes the electrons with the highest energy, $E_{\mathrm{e,max}}$, in a loss-limited SNR (see also Vink, 2004).

From Eqs.~(\ref{eq:DRadvNum}) and~(\ref{eq:DRdiffNum}), one easily obtains the thickness ratio in the case of a Bohm diffusion coefficient (at the observed gamma-ray energy of 5~keV, assumed to be close to $E_{\mathrm{e},\max}$):
\begin{equation}
\rho_{\mathrm{B}}\equiv \frac{\Delta R_{\mathrm{diff}}^{(\mathrm{Bohm})}}{\Delta R_{\mathrm{adv}}} \simeq 5.7 \times \frac{r}{4} \times V_{\mathrm{sh},3}^{-1},
\label{eq:DRdiffSurDRadv}
\end{equation}
independent of the magnetic field. The numerical values obtained with our set of parameters are shown in Table~\ref{tab:someLimits}, for $r = 4$. Given the uncertainties in the measured quantities, they are remarkably close to 1! This indicates that the value of the electron diffusion coefficient at $E_{\mathrm{e,max}}$ cannot be significantly different from the assumed Bohm value. In the next section, we make this argument more precise by investigating the high-energy cutoff of the acceleration process, taking into account the time spent by the particles both upstream and downstream, and without restriction to the Bohm diffusion regime.

\section{Constraints on the SN acceleration parameters}
\label{sec:diffusionRegime}

\subsection{Acceleration timescale}

Apart from a few ``lucky'' particles that gain energy on a shorter timescale than the average in the stochastic acceleration process, the maximum energy, $E_{\mathrm{e,max}}$, that an electron can reach is obtained by equating the synchrotron loss time to the acceleration time scale. According to the DSA theory, the latter is given by (Drury, 1983):
\begin{equation}
\tau_{\mathrm{acc}} = \frac{3r}{r-1}\left(\frac{rD_{\mathrm{d}}+D_{\mathrm{u}}}{V_{\mathrm{sh}}^2}\right),
\label{eq:tauAcc}
\end{equation}
where $D_{\mathrm{u}}$ and $D_{\mathrm{d}}$ are the upstream and downstream diffusion coefficients. 

Without any loss of generality, we can write $D_{\mathrm{u}} \equiv k_{\mathrm{u}} D_{\mathrm{B,u}}$ and $D_{\mathrm{d}} \equiv k_{\mathrm{d}} D_{\mathrm{B,d}}$, where $D_{\mathrm{B},i}$ is the Bohm diffusion coefficient in the local magnetic field (upstream or downstream), as given by Eq.~(\ref{eq:DBohm}), and $k_{i}$ is merely a number characterizing the difference between $D(E_{\mathrm{e,max}})$ and the Bohm value: $k_{i} \ge 1$. Note that in principle it depends on $E$, as in the case of Kolmogorov or Kraichnan type of turbulence, but this is not important here since we consider only the highest energy electrons.

In order to limit the number of free parameters, we make the reasonable assumption that the upstream and downstream magnetic fields are related according to the usual jump conditions (shock compression of the components parallel to the shock front). Given the high magnetic field values, which require field amplification, we assume that the magnetic turbulence is roughly isotropic upstream and write (following Berezhko, et al., 2002): $B_{\mathrm{d}} \simeq \sqrt{(1+2r^2)/3}B_{\mathrm{u}}\simeq 0.83rB_{\mathrm{u}}$. Taking the downstream magnetic field as a reference to compute the Bohm diffusion coefficient, one can rewrite Eq.~(\ref{eq:tauAcc}) as:
\begin{equation}
\tau_{\mathrm{acc}} = \frac{D_{\mathrm{B}}}{V_{\mathrm{sh}}^2}\frac{3r^2}{r - 1} (k_{\mathrm{d}} + 0.83 k_{\mathrm{u}}) \simeq 1.83 \,k_{0}\, \frac{D_{\mathrm{B}}}{V_{\mathrm{sh}}^2} \frac{3r^2}{r - 1},
\label{eq:tauAcc2}
\end{equation}
where we have assumed, in the last equality, that $k_{\mathrm{u}} \simeq k_{\mathrm{d}} \equiv k_{0}$, as expected if the amplified magnetic field has essentially the same structure upstream and downstream, apart from the above-mentioned shock compression. A more detailed investigation would involve additional free parameters describing the plasma properties on both sides of the shock and a complete model of turbulence, which is left for future work. Note, in particular, that a purely isotropic turbulence both upstream and downstream cannot hold in principle, because of the shock compression itself. It should also be noted that the effective value of the compression ratio ``felt'' by the particles depends in principle on their rigidity in the case of a strongly modified shock. Only the protons of highest energy should feel the total compression ratio, $r_{\mathrm{tot}}$ (including the precursor), while low-energy particles feel the much lower compression ratio of the sub-shock, $r_{\mathrm{sub}}$, and the electrons of highest energy explore a region across the shock with an intermediate effective compression ratio $r_{\mathrm{el}}$ (keeping in mind our basic assumption that the X-ray rim thickness is limited by synchrotron losses, so that these electrons have a lower energy than they would reach otherwise). We shall not go into such details here, as they would only complicate the formal treatment with no significant change in the results (e.g. always smaller than 30\% for the deduced value of $E_{\mathrm{p,max}}$; see below). In any case, we found that a situation with explicitly different values of the compression ratio for protons and electrons (and particles of different energies) in the case of a strongly modified shock results in an intermediate result between the test-particle case (no precursor nor sub-shock, and $r=4$) and the pure $r = 10$ case. We shall thus confine our study below to such idealised cases, using only one (``universal'') compression ratio for all particles. Numerically, Eq.~(\ref{eq:tauAcc2}) gives:
\begin{equation}
\tau_{\mathrm{acc}} \simeq (30.6\,\mathrm{yr}) \frac{3r^{2}}{16(r-1)}\times k_{0}(E)\times E_{\mathrm{TeV}}\, B_{100}^{-1}\,V_{\mathrm{sh},3}^{-2}.
\label{eq:tauAccNum}
\end{equation}

\begin{table*}
\begin{center}
\begin{tabular}{c|c|c|c|c|c|c|c|c|c|}
\multirow{2}*{SNR} & \multicolumn{2}{c|}{$k_{0}(E_{\mathrm{e,max}})$} & \multicolumn{4}{c|}{$B_{\mathrm{d}}(\alpha,r)$ in $\mu$G} & \multicolumn{2}{c|}{$E_{\mathrm{e,max}}(\alpha,r)$ in TeV} & $\alpha_{\mathrm{min}}$ (Eq.~\ref{eq:alphaMin}) \\
 & $r = 4$ & $r = 10$ & (1 ; 4) & (1 ; 10) & (1/3 ; 4) & (1/3 ; 10) & (1 ; 4) & (1/3 ; 10) & $r = 10$ \\
\hline
Cas A & 3.2 & 1.5 & 390 & 280 & 350 & 250 & 12 & 15 & 0.08 \\
Kepler & 4.5 & 2.2 & 340 & 250 & 300 & 210 & 11 & 14 & 0.09 \\
Tycho & 10 & 4.9 & 530 & 400 & 400 & 300 & 5.2 & 6.9 & 0.10 \\
SN 1006 & (0.40) & (0.19) & 110 (84) & 95 (59) & 100 (82) & 91 (57) & 37 (32) & $-$ & 0.37 \\
G347.3$-$0.5 & (0.87) & (0.41) & 96 (93) & 84 (66) & 92 (89) & 79 ( 62) & 36 (37) & $-$ & 0.13 \\
\end{tabular}
\end{center}
\caption{Diffusion and acceleration parameters estimated for the five young SNRs considered in Table~\ref{nt_snrs} (see text).}
\label{tab:SNRValues}
\end{table*}

\subsection{X-ray cut-off and diffusion coefficient}

Likewise, to evaluate the effective synchrotron loss time, we need to distinguish the upstream and downstream energy losses (cf. Yamazaki et al., 2004). If $\tau_{\mathrm{u}}$ and $\tau_{\mathrm{d}}$ are the time respectively spent upstream and downstream, and $\tau_{\mathrm{syn,u}}$ and $\tau_{\mathrm{syn,d}}$ the corresponding synchrotron time scales, the effective energy decay law over one cycle, $\tau_{\mathrm{cycle}} = \tau_{\mathrm{u}} + \tau_{\mathrm{d}}$, can be written $E(\tau_{\mathrm{cycle}}) = E_{0} \exp( - \tau_{\mathrm{u}}/\tau_{\mathrm{syn,u}}) \exp( - \tau_{\mathrm{d}}/\tau_{\mathrm{syn,d}})$. The average synchrotron loss time, $\left<\tau_{\mathrm{syn}}\right>$, is obtained by identification with $E(\tau_{\mathrm{cycle}}) = E_{0} \exp( - \tau_{\mathrm{cycle}}/\left<\tau_{\mathrm{syn}}\right>)$:
\begin{equation}
\left<\tau_{\mathrm{syn}}\right> = (1.25\,10^{3}\,\mathrm{yr})\times E_{\mathrm{TeV}}^{-1}\times \left<B_{100}^{2}\right>^{-1},
\label{eq:tauSynEff}
\end{equation}
where $\left<B^{2}\right> = (B_{\mathrm{u}}^{2}\tau_{\mathrm{u}} + B_{\mathrm{d}}^{2}\tau_{\mathrm{d}})/(\tau_{\mathrm{u}} + \tau_{\mathrm{d}})$ is the average square field. With $\tau_{i}\propto D_{i}/V_{i}$, we have $\tau_{\mathrm{u}}/\tau_{\mathrm{d}} = D_{\mathrm{u}}/rD_{\mathrm{d}} \simeq B_{\mathrm{d}}/rB_{\mathrm{u}}\simeq 0.83$ and
\begin{equation}
\left<B^{2}\right> \simeq B_{\mathrm{d}}^{2} \frac{1+1/0.83r^{2}}{1.83} \simeq 0.59\,B_{\mathrm{d}}^{2}.
\label{eq:BMeanApprox}
\end{equation}
Thus, in terms of the downstream magnetic field:
\begin{equation}
\left<\tau_{\mathrm{syn}}\right> \simeq (2.1\times 10^{3}\,\mathrm{yr})\times E_{\mathrm{TeV}}^{-1}\, B_{100}^{-2}.
\label{eq:tauSynchEgammaBis}
\end{equation}

From Eqs.~(\ref{eq:tauAccNum}) and~(\ref{eq:tauSynchEgammaBis}), one obtains:
\begin{equation}
E_{\mathrm{e,max}} \simeq (8.3\,\mathrm{TeV})\times \bar{f}(r)\times k_{0}^{-1/2}\times B_{100}^{-1/2}\,V_{\mathrm{sh},3},
\label{eq:EeMax}
\end{equation}
where $\bar{f}(r) \equiv f(r)/f(4)$, with $f(r) = \sqrt{r-1}/r$: $\bar{f}(r)$ takes values between 1 and $f(10)/f(4) \simeq 0.693$.

On the other hand, Eq.~(\ref{eq:EEgamma}) gives $E_{\mathrm{e,max}}$ as a function of the X-ray cutoff energy:
\begin{equation}
E_{\mathrm{e,max}} \simeq (22\,\mathrm{TeV})\times B_{100}^{-1/2}\times E_{\gamma,\mathrm{cut,keV}}^{1/2},
\label{eq:EeMaxBis}
\end{equation}
with the same dependence in $B$. Identifying both expressions, we can thus derive $k_{0} = D(E_{\mathrm{e,max}})/D_{\mathrm{B}}(E_{\mathrm{e,max}})$ directly from the SNR data, namely the X-ray cutoff energy and the shock velocity:
\begin{equation}
k_{0}(E_{\mathrm{e,max}}) = 0.14 \times E_{\gamma,\mathrm{cut,keV}}^{-1} \times V_{\mathrm{sh},3}^{2} \times \bar{f}(r)^{2}.
\label{eq:k0EeMax}
\end{equation}
with $0.48 \le \bar{f}(r)^{2} \le 1$. The values obtained for $r = 4$ and $r = 10$ are given in Table~\ref{tab:SNRValues}. $k_{0}$ is found to be larger than 1 for Cas~A, Kepler and Tycho, as required, or marginally lower than 1 for G347.3-0.5. In the case of SN~1006, $k_{0} \simeq 0.4$ for $r = 4$ and 0.2 for $r = 10$, which may favour low compression ratios and/or point to a possible overestimate of $E_{\gamma,\mathrm{max}}$ (e.g. if the synchrotron cutoff is less sharp than exponential) and/or underestimate of $V_{\mathrm{sh}}$. Note that the cut-off energy and shock velocity given in Table~\ref{nt_snrs} were obtained from different regions in the SNR. The cut-off energies are quite uncertain, being mostly obtained by comparing X-ray and radio fluxes. They can be overestimated if the spectrum is concave (predicted by non-linear acceleration), or underestimated if the radio flux is taken from a region larger than just the rim. Additional data on these parameters would be very valuable for the present study.

Despite the above uncertainties, it can be seen from Table~\ref{tab:SNRValues} that $k_{0}$ keeps reasonably small values, which means that the actual diffusion coefficient at $E_{\mathrm{e,max}}$ is not very much larger than the Bohm value, and possibly very close to it in the case of SN 1006 and G347.3-0.5. Note also that numerical studies of diffusion in a turbulent magnetic field found values of $k_{0} \simeq 3$--4 at the energy where $r_{\mathrm{L}} = \lambda_{\mathrm{c}}$, the coherent length of a turbulent magnetic field \citep{Cetal02,Parizot04}. If the parameter values for Cas~A and Kepler are confirmed, this would favour a turbulent scale close to $r_{\mathrm{L}}(E_{\mathrm{e,max}})$.

\subsection{Rim thickness and diffusion regime}
\label{sec:diffRegime}

We now repeat the calculations of Sect.~\ref{sec:constraints}, but taking into account advection and diffusion jointly, and using the above determination of the diffusion coefficient at $E_{\mathrm{e,max}}$. To derive the scale of the high energy electron distribution downstream, we write again the stationary version of Eq.~(\ref{eq:transportEquation}), keeping advective and diffusive terms: $V_{\mathrm{d}}(\partial f/\partial x) = D\partial^{2}f/\partial x^{2} - f/\tau_{\mathrm{syn}}$. The solution is again of the form $f(x)\propto \exp(-ax)$, where $a$ is the positive solution of the quadratic characteristic equation (Berezhko \& V\"olk, 2004):
\begin{equation}
Da^{2} + V_{\mathrm{d}}a - \frac{1}{\tau_{\mathrm{syn}}} = 0 \Rightarrow a = \sqrt{\frac{V_{\mathrm{d}}^{2}}{4D^{2}} + \frac{1}{D\tau_{\mathrm{syn}}}} - \frac{V_{\mathrm{d}}}{2D}
\label{eq:characteristicEquation}
\end{equation}
The scale of the emitting region is then $\Delta R_{\mathrm{rim}} = a^{-1}$, and the observed (projected) size of the rims thus reads:
\begin{equation}
\Delta R_{\mathrm{obs}} = P \times \frac{2D/V_{\mathrm{d}}}{\sqrt{1 + 4D/V_{\mathrm{d}}^{2}\tau_{\mathrm{syn}}} - 1}.
\label{eq:DRObs}
\end{equation}
By definition, this equation holds only at $E_{\mathrm{e,obs}}$, given by Eq.~(\ref{eq:EEgammaObs}), and we thus need to evaluate the synchrotron loss time and the diffusion coefficient at \emph{that} energy. We already obtained $\tau_{\mathrm{syn}}(E_{\mathrm{e,obs}})$ in Eq.~(\ref{eq:tauSynchEObs}). For the diffusion coefficient, we derived its value at $E_{\mathrm{e,max}}$ and now need make an assumption about the \emph{diffusion regime}, i.e. the energy dependence of $D$. This is not known a priori, so we shall leave it unspecified and simply assume a power-law form:
\begin{equation}
D(E) = D(E_{0})\left(\frac{E}{E_{0}}\right)^{\alpha} \hspace{-3pt} = k_{0} D_{\mathrm{B}}(E_{\mathrm{e,max}}) \left(\frac{E}{E_{\mathrm{e,max}}}\right)^{\alpha},
\label{eq:DofE}
\end{equation}
where the index $\alpha$ is a free parameter. The Bohm regime corresponds to $\alpha = 1$. Quasi-linear theory predicts $\alpha = 1/3$ (resp. 1/2) in the case of a Kolmogorov-like (resp. Kraichnan) spectrum of magnetic turbulence, and a simple calculation leads to $\alpha = 2$ for particles with a gyroradius larger than the coherence length of the field. Both of these behaviors are very well reproduced with numerical simulations of particle diffusion (e.g. Parizot, 2004).
Note that since $D(E)$ must remain larger than $D_{\mathrm{B}}(E)$, one must have $k_{0}(E/E_{0})^{\alpha-1} > 1$ at all energies where the power-law regime holds. Since $k_{0} \la 10$ at $E_{\mathrm{e,max}}$ and since the injection energy in the acceleration process is many orders of magnitude below $E_{\mathrm{e,max}}$, this implies that $\alpha \le 1$ at least up to energies close to $E_{\mathrm{e,max}}$.

Although neither $E_{\mathrm{e,obs}}$ nor $E_{\mathrm{e,max}}$ are known yet, their ratio has a very simple expression. The typical synchrotron photon energy is simply proportional to the electron energy squared, so by definition, with $E_{\gamma,\mathrm{obs}} = 5$~keV:
\begin{equation}
\frac{E_{\mathrm{e,obs}}}{E_{\mathrm{e,max}}} = \left(\frac{E_{\gamma,\mathrm{obs}}}{E_{\gamma,\mathrm{cut}}}\right)^{1/2} \simeq 2.2 \times E_{\gamma,\mathrm{cut,keV}}^{-1/2}.
\label{eq:EObsOverEMax}
\end{equation}
Since $E_{\gamma,\mathrm{cut}} < 5$~keV for all five SNRs, we have $E_{\mathrm{e,obs}} > E_{\mathrm{e,max}}$ (the ratios are respectively 2.0, 2.3, 4.1, 1.3 and 1.4).  This can be used to set a lower limit to $\alpha$, by requiring that $D(E_{\mathrm{e,obs}}) \ge D_{\mathrm{Bohm}}(E_{\mathrm{e,obs}})$. From Eq.~(\ref{eq:DofE}), this translates into:
\begin{equation}
\alpha \ge 1 - \frac{\ln k_{0}(E_{\mathrm{e,max}})}{\ln(E_{\mathrm{e,obs}}/E_{\mathrm{e,max}})}.
\label{eq:alphaMin2}
\end{equation}
This is however not constraining in the case of Cas~A, Kepler and Tycho, since the lower limit is negative. For SN 1006 and G347.3-0.5, we already noted that the value of $k_{0}$ derived from the data is lower than 1. In the following, we shall assume $k_{0} = 1$ in this case, putting the unrealistic computed value down to reasonable uncertainties in the measured parameters.

\subsection{Self-consistent magnetic field and maximum electron energy}

Replacing Eq.æ(\ref{eq:EObsOverEMax}) in~(\ref{eq:DofE}) and using~(\ref{eq:k0EeMax}), we get:
\begin{equation}
D(E_{\mathrm{e,obs}}) = (1.0\,10^{24}\,\mathrm{cm}^{2}\mathrm{s}^{-1}) B_{100}^{-3/2} V_{\mathrm{sh},3}^{2}  K(\alpha,r),
\label{eq:DEeObs}
\end{equation}
where we defined
\begin{equation}
K(\alpha,r) = \bar{f}(r)^{2} \times 2.2^{\alpha} \times E_{\gamma,\mathrm{cut,keV}}^{-1/2-\alpha/2}
\label{eq:K}
\end{equation}
Using Eq.~(\ref{eq:tauSynchEObs}), the ratio in the denominator of Eq.~(\ref{eq:DRObs}) follows:
\begin{equation}
\frac{4D(E_{\mathrm{e,obs}})}{V_{\mathrm{d}}^{2}\tau_{\mathrm{syn}}(E_{\mathrm{e,obs}})} = 8.1\left(\frac{r}{4}\right)^{2}K(\alpha,r),
\label{eq:denominator}
\end{equation}
which is independent of both $B_{\mathrm{d}}$ and $V_{\mathrm{sh}}$. Replacing Eqs.~(\ref{eq:denominator}) and~(\ref{eq:DEeObs}) in~(\ref{eq:DRObs}), we obtain a self-consistent expression of the downstream magnetic field as a function of $\alpha$, $r$ and the measured SNR parameters:
\begin{equation}
B_{\mathrm{d}} \simeq (520\,\mu\mathrm{G}) \left[\frac{(r/4)\bar{P}K(\alpha,r)V_{\mathrm{sh,3}} \Delta R_{\mathrm{obs,-2}}^{-1}}{\sqrt{1 + 8.1(r/4)^{2}K(\alpha,r)} - 1}\right]^{2/3}.
\label{eq:Bd}
\end{equation}
The results are given in Table~\ref{tab:SNRValues}, for different compression ratios and diffusion regimes. For SN 1006 and G347.3-0.5, we assumed $k_{0}(E_{\mathrm{e,max}}) = 1$ (i.e. $D = D_{\mathrm{Bohm}}$) and gave the value obtained with the (non physical) computed value of $k_{0} < 1$ between parentheses. As already noted, the estimated magnetic fields are much larger than those typical of the interstellar medium and require strong amplification at the shock. Lower magnetic fields are obtained with lower values of $\alpha$ and/or larger compression ratios. Two limiting cases are considered: $(\alpha = 1, r = 4)$ and ($\alpha = 1/3, r = 10)$.

\begin{figure*}[t]
\centering
\includegraphics[width=0.49\linewidth]{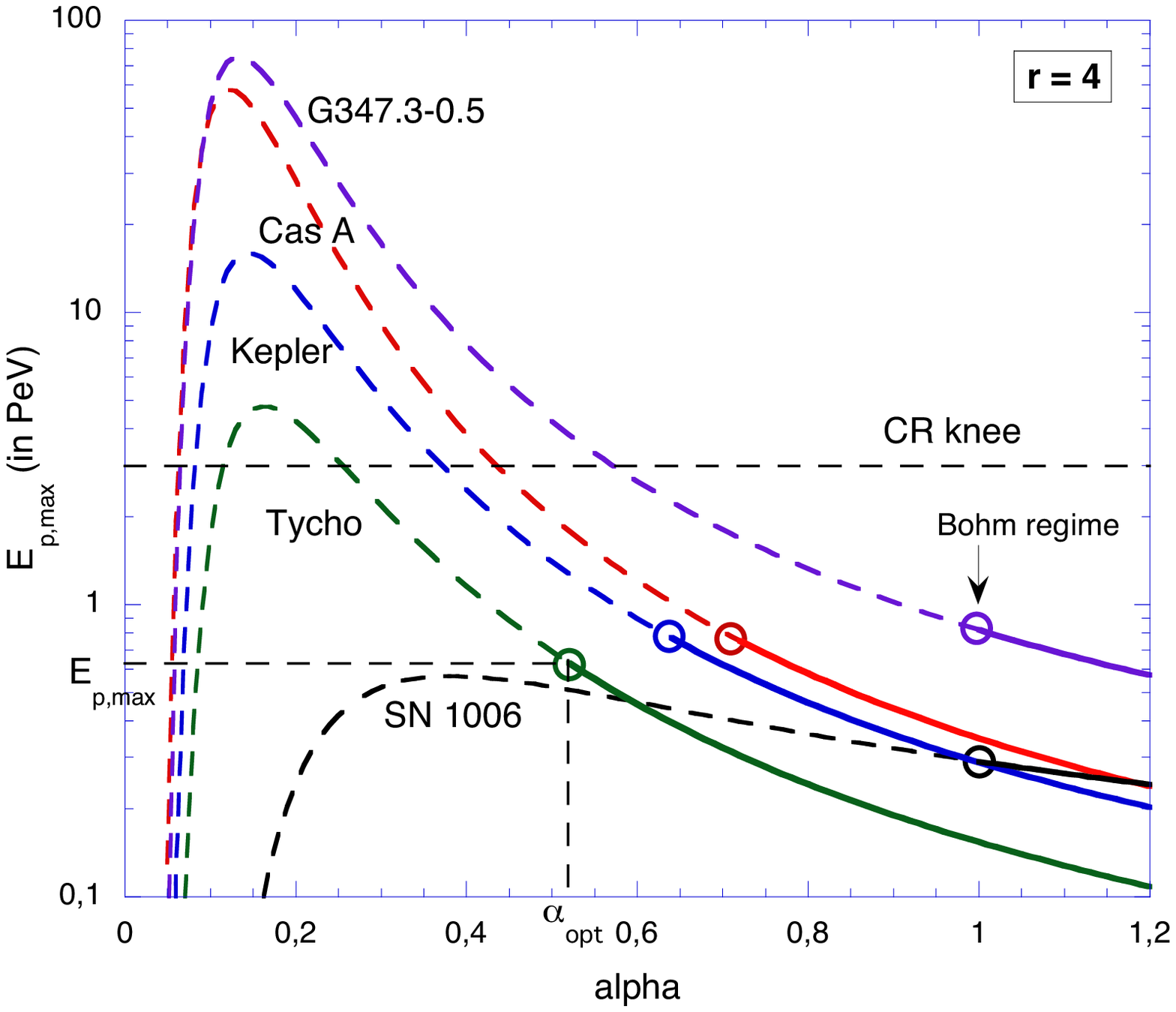}\hfill \includegraphics[width=0.49\linewidth]{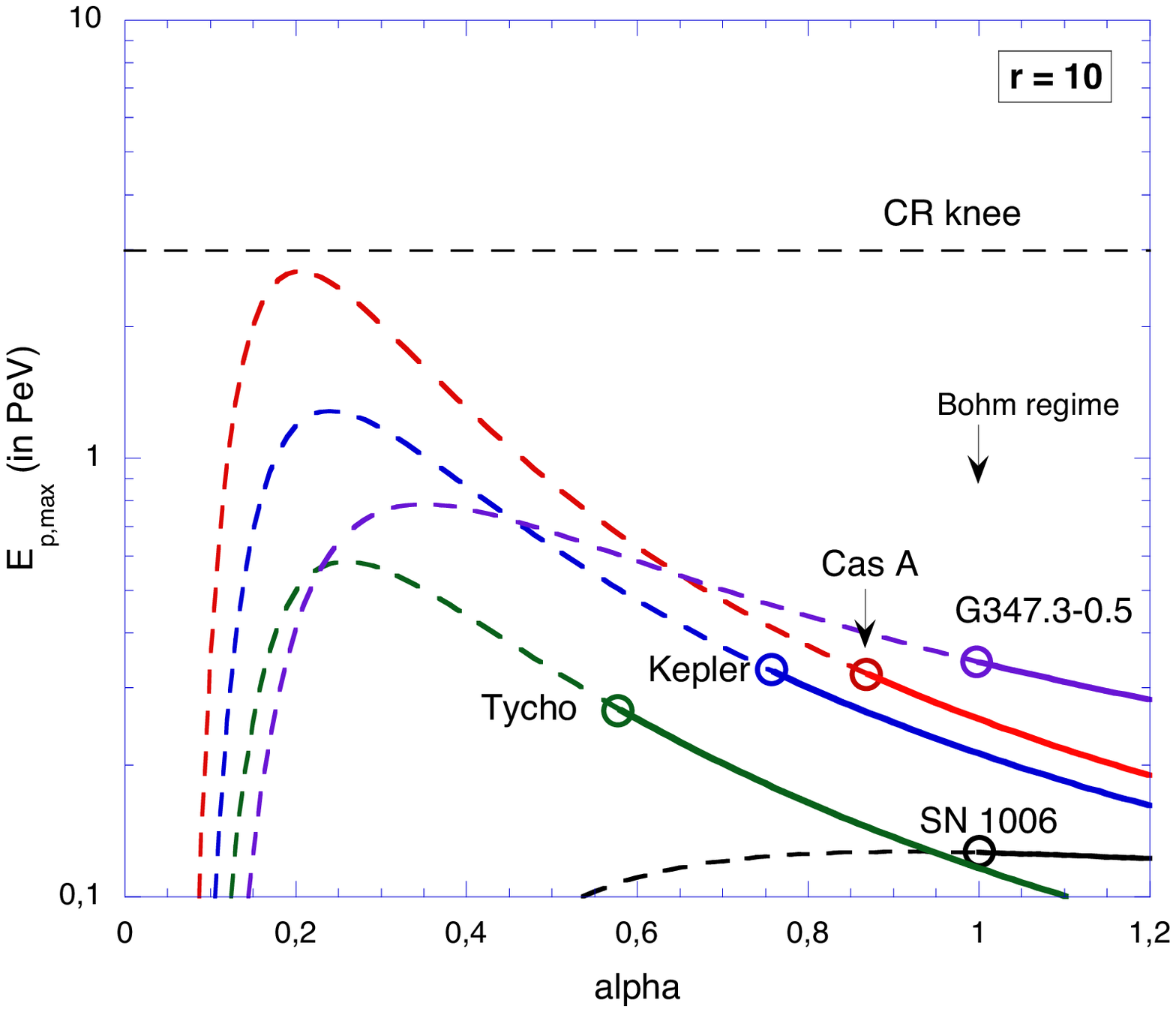}
\caption{Maximum proton energy, $E_{\mathrm{p,max}}$, as a function of the diffusion regime index $\alpha$, for $r = 4$ (left) and $r = 10$ (right). The dashed part of the curves correspond to values of $\alpha$ that lead to a diffusion coefficient smaller than the Bohm limit at $E_{\mathrm{p,max}}$, and are therefore excluded. The highest possible value of $E_{\mathrm{p,max}}$ is materialised by the circles giving the value $\alpha_{\mathrm{opt}}$ above which diffusion regime indices are valid -- notably the Bohm index $\alpha = 1$, but lead to smaller values of $E_{\mathrm{p,max}}$.}.
\label{fig:EpMax}
\end{figure*}

With the above value of the magnetic field, we can now derive the maximum electron energy in a fully consistent way, from Eq.~(\ref{eq:EeMaxBis}):
\begin{equation}
\begin{split}
E_{\mathrm{e,max}} \simeq \, &(9.6\,\mathrm{TeV}) \times E_{\gamma,\mathrm{cut,keV}}^{1/2}\\ &\times \left[\frac{(r/4)\bar{P}K(\alpha,r)V_{\mathrm{sh,3}} \Delta R_{\mathrm{obs,-2}}^{-1}}{\sqrt{1 + 8.1(r/4)^{2}K(\alpha,r)} - 1}\right]^{-1/3}.
\label{eq:EeMaxFromSNRParam}
\end{split}
\end{equation}
The results are given in Table~\ref{tab:SNRValues}: they are of the order of 10~TeV for Cas A and Kepler, 5~TeV for Tycho, and 40~TeV for SN 1006 and G347.3-0.5, almost insensitive to the parameters $\alpha$ and $r$ (the two extreme cases are shown).

Finally, the value of $E_{\mathrm{e,max}}$ and the associated value of $E_{\mathrm{e,obs}}$ allow us to set a lower limit to $\alpha$, by requiring that the total acceleration time of electrons up to the highest energies do not exceed the age of the SNR, $t_{\mathrm{SNR}}$. Integrating the acceleration timescale, Eq.~(\ref{eq:tauAcc}), from $E_{\mathrm{inj}}$ to $E_{\mathrm{e,obs}} \gg E_{\mathrm{inj}}$, using Eq.~(\ref{eq:DofE}), we find (for $\alpha > 0$)
\begin{equation}
t_{\mathrm{acc}}(E_{\mathrm{e,obs}}) = \int_{E_{\mathrm{inj}}}^{E_{\mathrm{e,obs}}} \tau_{\mathrm{acc}}(E)\frac{\mathrm{d}E}{E} \simeq \frac{1}{\alpha}\,\tau_{\mathrm{acc}}(E_{\mathrm{e,obs}}).
\label{eq:tAcc}
\end{equation}
Now from Eqs.~(\ref{eq:tauAcc2}) and~(\ref{eq:DEeObs}), we get:
\begin{equation}
\begin{split}
\tau_{\mathrm{acc}}(E_{\mathrm{e,obs}}) &\simeq (93\,\mathrm{yr})\bar{f}^{-2} B_{\mathrm{100}}^{-3/2} K(\alpha,r)\\
&\simeq (93\,\mathrm{yr})B_{\mathrm{100}}^{-3/2} \times 2.2^{\alpha} \,E_{\gamma,\mathrm{cut,keV}}^{-1/2-\alpha/2},
\label{eq:tauAccEeObs}
\end{split}
\end{equation}
so that the condition $t_{\mathrm{acc}}(E_{\mathrm{e,obs}}) < t_{\mathrm{SNR}}$ reads:
\begin{equation}
\frac{\alpha\,\bar{f}^{2}\bar{P}(r/4)}{\sqrt{1 + 8.1(r/4)^{2}K(\alpha,r)} - 1} \ga \frac{7.8\,\mathrm{yr}}{t_{\mathrm{SNR}}}\,\frac{\Delta R_{\mathrm{obs},-2}}{V_{\mathrm{sh},3}},
\label{eq:alphaMin}
\end{equation}
Solving for $\alpha$, we get the lower limit $\alpha_{\mathrm{min}}$. For $r = 4$, this is of the order of $0.05$, which is not really constraining. The values of $\alpha_{\mathrm{min}}$ are given in Table~\ref{tab:SNRValues} for $r = 10$. They also are not constraining, and compatible with either a Kolmogorov-like or a Bohm diffusion regime (note that the case $r = 10$ with $\alpha < 1$ is not relevant for SN~1006 and G347.3-0.5 anyway).

\begin{table*}
\begin{center}
\begin{tabular}{|c|c|c|c|c|c|c|c|c|c|}
\hline
\multirow{3}*{SNRs} & \multicolumn{3}{c|}{$r = 4$} & \multicolumn{3}{c|}{$r = 10$} & \multicolumn{3}{c|}{modified SNR parameters} \\
\cline{2-10}
 & $\alpha_{\mathrm{opt}}$ & \multicolumn{2}{c|}{$E_{\mathrm{p,max}}$ (in PeV)} & $\alpha_{\mathrm{opt}}$ & \multicolumn{2}{c|}{$E_{\mathrm{p,max}}$ (in PeV)} & $\alpha_{\mathrm{opt}}$ & \multicolumn{2}{c|}{$E_{\mathrm{p,max}}$ (in PeV)} \\
\cline{3-4} \cline{6-7} \cline{9-10}
 & ($= \alpha_{\min}$) & $\alpha = \alpha_{\mathrm{opt}}$ & $\alpha = 1$ & ($= \alpha_{\min}$) & $\alpha = \alpha_{\mathrm{opt}}$ & $\alpha = 1$ & ($= \alpha_{\min}$) & $\alpha = \alpha_{\mathrm{opt}}$ & $\alpha = 1$ \\
\hline
Cas A & 0.72  & 0.75 & 0.35 & 0.87 & 0.32 & 0.25 & 0.71 & 2.1 & 0.64 \\
Kepler & 0.64 & 0.77 & 0.29 & 0.76 & 0.33 & 0.21 & 0.65 & 2.2 & 0.53 \\
Tycho & 0.52 & 0.63 & 0.15 & 0.58 & 0.27 & 0.11 & 0.56 & 2.0 & 0.28 \\
SN 1006 & 1 & 0.29 & 0.29 & 1 & 0.13 & 0.13 & 1 & 0.73 & 0.73 \\
G347.3$-$0.5 & 1 & 0.82 & 0.82 & 1 & 0.34 & 0.34 & 0.94 & 2.0 & 1.7 \\
\hline
\end{tabular}
\end{center}
\caption{Derived values of $\alpha$ and $E_{\mathrm{p,max}}$ for the five SNRs considered in Table~\ref{nt_snrs} (see text). In addition to the cases of a compression ratio of 4 and 10, we consider an ``extreme case'' with SNR parameters artificially modified so as to increase the value of $E_{\mathrm{p,max}}$: this amounts to increasing the shock velocity by 25\% and dividing the observed rim thickness by a factor of 2. In each case, we give of $E_{\mathrm{p,max}}$ for $\alpha_{\mathrm{opt}}$ and for $\alpha = 1$ (i.e. Bohm diffusion regime).}
\label{tab:EPMax}
\end{table*}

\section{Maximum energy of the accelerated protons}

In the previous section, we estimated the value of the amplified magnetic field in the vicinity of the shock and of the diffusion coefficient at the highest electron energy, $E_{\mathrm{e,max}}$. We now consider their implications for the highest proton energy, $E_{\mathrm{p,max}}$, which (in the most favourable case) is limited by the age of the SNR and is thus obtained from the condition $t_{\mathrm{acc}}(E_{\mathrm{p,max}}) = t_{\mathrm{SNR}}$. Integrating like in Eq.~(\ref{eq:tAcc}) up to $E_{\mathrm{p,max}}$, one obtains:
\begin{equation}
t_{\mathrm{acc}}(E_{\mathrm{p,max}}) = \frac{\tau_{\mathrm{acc}}(E_{0})}{\alpha} \left(\frac{E_{\mathrm{p,max}}}{E_{0}}\right)^{\alpha},
\label{eq:tAccProtons}
\end{equation}
and thus
\begin{equation}
E_{\mathrm{p,max}} \simeq E_{0} \left[\frac{\alpha\, t_{\mathrm{SNR}}}{\tau_{\mathrm{acc}}(E_{0})}\right]^{1/\alpha}.
\label{eq:EpMaxFormal}
\end{equation}

Choosing $E_{0} = E_{\mathrm{e,max}}$ or $E_{\mathrm{e,obs}}$ and replacing the corresponding expressions, we derive $E_{\mathrm{p,max}}$ as a function of the only remaining free parameters, $\alpha$ and $r$ (through $B_{100}$):
\begin{equation}
E_{\mathrm{p,max}} \simeq (49\,\mathrm{TeV}) B_{100}^{-1/2} \left[\frac{t_{\mathrm{SNR}}}{93\,\mathrm{yr}}\,\frac{\alpha B_{100}^{3/2}}{2.2^{\alpha}} E_{\gamma,\mathrm{cut,keV}}^{1/2+\alpha/2}\right]^{1/\alpha},
\label{eq:EpMax}
\end{equation}
where $B$ is given by Eq.~(\ref{eq:Bd}).
This maximum proton energy is shown as a function of $\alpha$ for the SNRs under consideration in Fig.~\ref{fig:EpMax}. The shape of the curve is easily understood by considering the interplay between the instantaneous acceleration rate, $\tau_{acc}$, and the integrated acceleration time, $t_{\mathrm{acc}}$, at $E_{\mathrm{p,max}}$. Above, we were able to calculate the diffusion coefficient at $E_{\mathrm{e,max}}$, which gives a fulcrum for $D$ at all other energies. For very low values of $\alpha$, the diffusion coefficient at low energy is much higher than with the Bohm diffusion, and the integrated acceleration time becomes very long. As $\alpha$ increases, this situation gets better, but the ratio between $D(E_{\mathrm{p,max}})$ and $D(E_{\mathrm{e,max}})$ also increases, so that $\tau_{\mathrm{acc}}(E_{\mathrm{p,max}})$ becomes higher and it takes longer to reach higher energies. The best compromise is for $\alpha\simeq 0.1$--0.2. However, these values of $\alpha$ are excluded because they lead to diffusion coefficients at $E_{\mathrm{p,max}}$ that are much lower than the Bohm diffusion coefficient. Therefore, the curves in Fig.~\ref{fig:EpMax} are only valid for pairs of values of $\alpha$ and $E_{\mathrm{p,max}}$ such that $D(E_{\mathrm{p,max}}) \ge D_{\mathrm{Bohm}}(E_{\mathrm{p,max}})$. Exactly as in Sect.~\ref{sec:diffRegime}, we must make sure that $k_{0} (E_{\mathrm{p,max}}/E_{\mathrm{e,max}})^{\alpha - 1} \ge 1$, which corresponds to the part of the curves shown in plain line (dashed line otherwise). Therefore, although relatively high values of $E_{\mathrm{p,max}}$ would seem to be reached for $\alpha\simeq 0.1$--0.2, notably higher than the energy of the knee, these values are not physical and must be excluded. 

\begin{figure}[t]
\centering
\includegraphics[width=\linewidth]{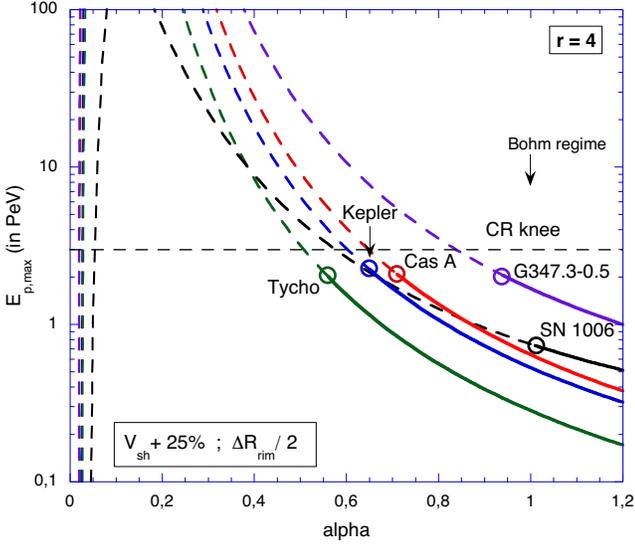}
\caption{Same as Fig.~\ref{fig:EpMax}a, with artificially modified SNR parameters: $V_{\mathrm{sh}}$ higher by 25\% and $\Delta R_{\mathrm{rim}}$ two times lower.}
\label{fig:EpMaxAllMoved}
\end{figure}

For each SNR, one may define the ``optimal value'' of $\alpha$ as the value $\alpha_{\mathrm{opt}}$ giving the highest possible value of $E_{\mathrm{p,max}}$. These correspond to the smallest ``allowed'' values of $\alpha$ as discussed above, and are given in Table~\ref{tab:EPMax}, together with the highest proton energy obtained under the assumption of a Bohm regime. As can be seen, $\alpha_{\mathrm{opt}}$ is typically between 0.5 and 0.7 for $r = 4$ and between 0.6 and 0.9 for $r = 10$, except for SN~1006 and G347.3-0.5, where we already noted that $\alpha < 1$ is not favoured. As can be seen, even in the optimal case, the highest possible proton energy falls short of the energy of the knee in the cosmic-ray spectrum, at $3\,10^{15}$~eV, by about a factor of 4 (or even 10 in the case of SN~1006). This is even more true if efficient acceleration is assumed, so that non-linear acceleration effects lead to a higher compression. For $r = 10$, $E_{\mathrm{p,max}}$ is systematically lower than the knee energy by one order of magnitude. It should also be noted that a Bohm diffusion \emph{regime} would give lower values of $E_{\mathrm{p,max}}$, which is due to the fact that $D(E_{\mathrm{p,max}})$ is then larger than the Bohm \emph{value}. The diffusion regime index that gives the lowest possible diffusion coefficient at $E_{\mathrm{p,max}}$ is precisely $\alpha_{\mathrm{opt}}$, hence the higher maximum energy.

From the above analyses alone, our results do not set any interesting constraint on the actual value of $\alpha$. Values lower than $\alpha_{\mathrm{opt}}$ may be considered as perfectly valid and merely resulting in lower values of the maximum proton energy. For instance, one might imagine a situation where electrons and protons have the same maximum energy, with the maximum turbulence scale fixed by $E_{\mathrm{e,max}}$ so that protons with a higher energy are not scattered resonantly and are thus not accelerated efficiently. However, we now claim that a further step can be taken if we consider the underlying turbulence generation process. While we just mentioned the possibility that the maximum scale of turbulence be set by electrons alone, so that the protons cannot be accelerated above $E_{\mathrm{e},\max}$, there is no reason \emph{a priori} why such a situation would occur. If the ultra-relativistic electrons are able to generate the turbulence necessary to keep the acceleration process going, then so should the protons, since at these energies they are essentially equivalent from the electromagnetic point of view. Besides, electrons only carry a negligible fraction of the energy of relativistic particles. It is thus reasonable to assume that whatever underlying processes actually lead protons to $E_{\mathrm{e},\max}$, they should not cease to work at that particular energy, which is determined by limiting processes affecting electrons only, nor at any other energy bearing no physical meaning for the process under consideration. In other words, the only natural limitation for the proton maximum energy is related to the acceleration time, $t_{\mathrm{acc}}(E_{\mathrm{p}}) \le t_{\mathrm{SNR}}$ (since the size-related limitation is less stringent for the SNRs considered here).

Applying this argument to the present study, we can actually reject any value of $\alpha$ lower than $\alpha_{\mathrm{opt}}$, since for these values the maximum proton energy is not limited by the age of the SNR, but ``artificially'' by imposing a cutoff in the turbulence spectrum at the energy where $D(E) = D_{\mathrm{Bohm}}(E)$. If that energy has to have a physical meaning, in terms of the Larmor radius of the highest energy protons (limited by the SNR age) in the amplified magnetic field, then $\alpha$ must be larger than the very value that we noted $\alpha_{\mathrm{opt}}$ above. From this point of view, $E_{\mathrm{p},\max}$ is \emph{known} to be given by Eq.~(\ref{eq:EpMaxFormal}), and the requirement that $D(E_{\mathrm{p},\max})$ be larger than $D_{\mathrm{Bohm}}(E_{\mathrm{p},\max})$, together with the observational constraint on $D(E_{\mathrm{e},\max})$, straightforwardly leads to $\alpha \ge \alpha_{\mathrm{opt}}$, i.e. to an interpretation of $\alpha_{\mathrm{opt}}$ as a lower limit for the diffusion coefficient index (assuming that the index does not significantly depend on energy, which may be expected to be the case up to energies close to the maximum resonant energy, such that $r_{\mathrm{L}}\simeq \lambda_{\max}$; Casse et al., 2002). Higher values of $\alpha$ remain possible in principle, since there is no reason \emph{a priori} to require that the diffusion coefficient at the highest proton energy (limited by $t_{SNR}$) have exactly the Bohm value.

Another interesting result is that $E_{\mathrm{p,max}}$ is roughly identical for all the SNRs considered, with values around 0.7 PeV at most, despite the different values of their parameters. This may indicate that not only these SNRs, but most of them probably cannot accelerate protons up to the knee. Since all the processes involved only depend on the rigidity of the nuclei, heavier nuclei of charge $Z$ should reach energies $Z$ times larger, i.e. $\la 20$~PeV for Fe nuclei in the optimal case.

In order to investigate the errors involved in our calculation due to uncertainties on the measured parameters of the SNRs, we considered what may be thought of as an extreme case, where the parameters have been moved consistently towards values leading to higher maximum proton energies. In Fig.~\ref{fig:EpMaxAllMoved}, we show $E_{\mathrm{p,max}}(\alpha)$ for this case where we have increased the shock velocity by 25\% and divided by 2 the rim thickness, keeping the most favourable value of the compression ratio, namely $r = 4$. The influence of the X-ray cut-off energy is somewhat more complicated, since a higher value of $E_{\gamma,\mathrm{cut}}$ results in an upper shift of the curves in Fig.~\ref{fig:EpMax}, thereby increasing $E_{\mathrm{p},\max}$ at a given $\alpha$, but at the same time the allowable range of values of $\alpha$ is reduced, and $\alpha_{\mathrm{opt}}$ is shifted upwards, reducing the maximum $E_{\mathrm{p},\max}$. The net effects of an upper (resp. lower) shift of $E_{\gamma,\mathrm{cut}}$ by a factor of 2 is then an increase (resp. decrease) of $E_{\mathrm{p},\max}$ by less than 10\%. Since the adopted values of $E_{\gamma,\mathrm{cut}}$ may already be considered upper limits (see above), we did not change this parameter in Fig.~\ref{fig:EpMaxAllMoved}. The corresponding values of $E_{\mathrm{p,max}}$ for $\alpha_{\mathrm{opt}}$ and for the Bohm diffusion regime are given in Table~\ref{tab:EPMax}. As can be seen, even in this case $E_{\mathrm{p,max}}(\alpha_{\mathrm{opt}})$ remains consistently below the knee, although only by 40\% for all SNRs except SN~1006. For $\alpha = 1$, $E_{\mathrm{p,max}}$ is still a factor of 4--10 below the knee for all SNRs except G347.3-0.5, where the factor is $\sim 2$. In other words, it appears that the strong magnetic field amplification reported here is still not enough to enable young, isolated SNRs as considered here to provide (at least directly) the main contribution to the observed Galactic cosmic-rays above $10^{15}$~eV, i.e. in the knee region and beyond.

\section{Conclusion}

In this paper, we have investigated which constraints on the parameters of young SNRs could be derived from the observed thickness of their X-ray rims, under the assumption that this thickness is limited by the synchrotron losses of the highest energy electrons -- which is fully consistent with our results. This extends the analysis of Berezhko \& V\"olk (2004) simply by relaxing the assumption of Bohm diffusion, and then provides an opportunity to constrain the diffusion coefficients from the data, in addition to other SNR parameters. Considering both the diffusion and advection of electrons in the downstream medium, we have shown that the magnetic field must be amplified in all the SNRs considered, with downstream values between 250 and 500~$\mu$G in the case of Cas A, Kepler, and Tycho, assuming a shock compression ratio of 4 (unmodified strong shock). If one assumes a compression ratio as high as~10, as may be the case if substantial shock modification occurs (very efficient shock acceleration), magnetic field values about 25\% lower are obtained, still requiring strong field amplification at the shock.

The value of the diffusion coefficient at the highest electron energy, $E_{\mathrm{e,max}}$, can also be derived from the data, by relating the X-ray energy cutoff to the acceleration timescale. We found that $D(E_{\mathrm{e,max}})$ is typically between 1 and 10 times the Bohm diffusion coefficient. We were also able to derive the value of $E_{\mathrm{e,max}}$, which ranges between 5 and 50~TeV, roughly independently of the compression ratio.

This analysis essentially constrains the diffusion coefficient at $E_{\mathrm{e,max}}$ and provides no significant information about the diffusion regime, i.e. the dependence of $D$ on energy. The comparison of the electron acceleration timescale with the age of the SNR showed that the data is compatible with any value of the energy dependence power-law index in the most interesting range: $1/3 \le \alpha \le 1$. However, by considering in the same framework the acceleration of protons and notably the maximum proton energy, we were able to define an ``optimal value'' of $\alpha$, typically in the range 0.5--1., for which the highest proton maximum energy could be reached. Further noting that the only natural limitation to $E_{\mathrm{p},\max}$ is the age of the SNR (in the cases under consideration), we also argued that the above value of $\alpha_{\mathrm{opt}}$ actually sets a lower limit to the possible diffusion coefficient index in these SNRs. In other words, values of $\alpha$ lower than $\alpha_{\mathrm{opt}}$ would lead to diffusion coefficients lower than the Bohm value at $E_{\mathrm{p},\max}$, as determined by $t_{\mathrm{acc}}(E_{\mathrm{p},\max}) = t_{\mathrm{SNR}}$. This favours diffusion regimes between the Kraichnan regime ($\alpha = 1/2$) and the Bohm regime ($\alpha = 1$), rather than a Kolmogorov-like regime, with $\alpha = 1/3$.

Finally, we deduced the maximum energy of nuclei accelerated in the five young SNRs under study. Although this energy depends on the diffusion regime, a general upper limit could be derived. We found $E_{\mathrm{max}} \simeq Z\times 0.7$--0.8~PeV for all SNRs (except for SN 1006: 0.3 PeV), obtained with the optimal value of $\alpha \sim 0.5$--0.7 (or 1 for SN 1006 and G347.3-0.5). Assuming a Bohm diffusion regime would lead to lower values of $E_{\mathrm{p,max}}$, around 0.3~PeV. Likewise, a higher compression ratio ($r = 10$) would correspond to $E_{\mathrm{max}}\la Z\times 0.3$~PeV in the optimal case.

Of course, the values calculated here depend on the accuracy of the observed SNR parameters, namely the synchrotron cut-off energy, the velocity of the expanding shock and the size of the X-ray rims. We investigated the influence of a misestimate of these parameters (the most influential of which is $E_{\gamma,\mathrm{cut}}$) by quite large amounts and found only small modifications of the highest proton energy, remaining marginally, but consistently lower than the knee energy. It should be remembered, however, that all our results were obtained under the assumption of a uniform, isotropic turbulence. We nevertheless point out that non uniform turbulence with the thickness of the rims limited by the relaxation scale of the turbulence (Pohl et al. 2005) would actually relax the constraint on the downstream magnetic field and make the acceleration less efficient, especially at large gyroradii, leading to even lower values of $E_{\mathrm{p,max}}$. 

In conclusion, the derived limit on the maximum energy of protons and nuclei accelerated in the young SNRs considered here appears to be quite robust, of the order of 800~TeV or up to 2~PeV by combining all uncertainties towards that direction, even though the magnetic field has been considerably amplified at the shock. In the most popular framework where the Galactic component of cosmic-rays extends up to the ankle, i.e. up to $\sim 3000$~PeV, our results would thus suggest that if isolated SNRs are the main sources of GCRs, an additional CR component is required above the knee(s). More specifically, it appears very difficult for the considered SNRs in their current stage of evolution to produce protons up to the knee of the cosmic-ray spectrum, at $\sim 3\,10^{15}$~eV, and essentially impossible to accelerate Fe nuclei up to either the ankle or the second knee at $\sim 500$~PeV. However, if the knee is associated with the contribution of a single nearby supernova (Erlykin \& Wolfendale 1997, 2004) dominating the bulk of the Galactic cosmic-rays produced by some other mechanism, the expected contribution, as derived here directly from measured parameters, would show the successive cut-off of the most abundant nuclei according to $E_{max}(i) \simeq Z_{i}\times 0.5$--1~PeV.

\begin{acknowledgements}
This work was partly supported by the International Space Science Institute (Bern) through the International Team Program, and by the French program of GDR PCHE. EP and AM are grateful to Andrei Bykov for leading a team on supernova remnants and thank its members for enlightening discussions. AM thanks J. Vink for many fruitful discussions.

\end{acknowledgements}

\bibliographystyle{aa}

\end{document}